%% file: early-warning-paper.tex
\documentclass[preprint2]{aastex63}
\usepackage[nolist]{acronym}
\usepackage{placeins}
\usepackage[utf8]{inputenc}
\usepackage[caption=false]{subfig}

\newcommand{\NumNSBH}{\ensuremath{2}}
\newcommand{\NumBNS}{\ensuremath{2}}

\newcommand{\NumPublishedCandidates}{\ensuremath{49}}

\newcommand{\NumPublicAlerts}{\ensuremath{56}}

\newcommand{\NumGstLALEWUploads}{$82$}
\newcommand{\NumSPIIREWUploads}{$141$}

\newcommand{\NumEWRetractions}{$5$}
\newcommand{\NumGstLALEWRetractions}{$4$}
\newcommand{\NumSPIIREWRetractions}{$1$}
\newcommand{\NumFullBandCircs}{$3$}

\newcommand{\EWFARThresh}{1 per day}
\newcommand{\FARThresh}{1 per 2 months}

\newcommand{\gwcelery}{GWCelery}

\newcommand{\gracedb}{GraceDB}
\newcommand{\lvalert}{LVAlert}

\newcommand{\SimMinMass}{$1.0\thinspace M_\odot$}
\newcommand{\SimMaxMass}{$2.0\thinspace M_\odot$}
\newcommand{\SimMeanMass}{$1.33\thinspace M_\odot$}
\newcommand{\SimStdev}{$0.09\thinspace M_\odot$}
\newcommand{\SimMaxRedshift}{0.2}

\newcommand{\mpc}{\ensuremath{\thinspace\text{Mpc}}}

\newif\ifshowparagraph
\showparagraphfalse 
\renewcommand{\paragraph}[1]{\ifshowparagraph #1 \fi}

\shorttitle{First demonstration of early warning gravitational-wave alerts}
\shortauthors{Magee, Chatterjee, Singer, Sachdev, et al.}

\begin{document}

\title{First demonstration of early warning gravitational wave alerts}

\author[0000-0001-9769-531X]{Ryan Magee}
\affiliation{LIGO, California Institute of Technology, Pasadena, CA 91125, USA}
\affiliation{Department of Physics, The Pennsylvania State University, University Park, PA 16802, USA}
\affiliation{Institute for Gravitation and the Cosmos, The Pennsylvania State University, University Park, PA 16802, USA}

\author[0000-0003-0038-5468]{Deep Chatterjee}
\affiliation{Center for Astrophysical Surveys, National Center for Supercomputing Applications, Urbana, IL, 61801, USA}
\affiliation{Illinois Center for Advanced Studies of the Universe, Department of Physics, University of Illinois at Urbana-Champaign, Urbana, IL 61801, USA}
\affiliation{Leonard E. Parker Center for Gravitation, Cosmology, and Astrophysics, University of Wisconsin-Milwaukee, Milwaukee, WI 53201, USA}

\author[0000-0001-9898-5597]{Leo P. Singer}
\affiliation{Astroparticle Physics Laboratory, NASA Goddard Space Flight Center, Mail Code 661, Greenbelt, MD 20771, USA}

\author[0000-0002-0525-2317]{Surabhi Sachdev}
\affiliation{Department of Physics, The Pennsylvania State University, University Park, PA 16802, USA}
\affiliation{Institute for Gravitation and the Cosmos, The Pennsylvania State University, University Park, PA 16802, USA}

\collaboration{4}{These authors contributed equally to this work}

\nocollaboration{32}

\author[0000-0001-8143-9696]{Manoj Kovalam}
\affiliation{Australian Research Council Centre of Excellence for Gravitational Wave Discovery (OzGrav)}
\affiliation{Department of Physics, University of Western Australia, Crawley WA 6009, Australia}

\author[0000-0001-6331-112X]{Geoffrey Mo}
\affiliation{LIGO Laboratory, Massachusetts Institute of Technology, Cambridge, MA 02139, USA}
\affiliation{Department of Physics and Kavli Institute for Astrophysics and Space Research,
Massachusetts Institute of Technology, Cambridge, MA 02139, USA}

\author{Stuart Anderson}
\affiliation{LIGO, California Institute of Technology, Pasadena, CA 91125, USA}

\author{Patrick Brady}
\affiliation{Leonard E. Parker Center for Gravitation, Cosmology, and Astrophysics, University of Wisconsin-Milwaukee, Milwaukee, WI 53201, USA}

\author{Patrick Brockill}
\affiliation{Leonard E. Parker Center for Gravitation, Cosmology, and Astrophysics, University of Wisconsin-Milwaukee, Milwaukee, WI 53201, USA}

\author{Kipp Cannon}
\affiliation{Research Center for the Early Universe, The University of Tokyo, 113-0033, Japan}

\author[0000-0001-5078-9044]{Tito Dal Canton}
\affiliation{Universit\'e Paris-Saclay, CNRS/IN2P3, IJCLab, 91405 Orsay, France}

\author{Qi Chu}
\affiliation{Australian Research Council Centre of Excellence for Gravitational Wave Discovery (OzGrav)}
\affiliation{Department of Physics, University of Western Australia, Crawley WA 6009, Australia}

\author{Patrick Clearwater}
\affiliation{Australian Research Council Centre of Excellence for Gravitational Wave Discovery (OzGrav)}
\affiliation{Gravitational Wave Data Centre, Swinburne University, Hawthorn VIC 3122, Australia}

\author{Alex Codoreanu}
\affiliation{Australian Research Council Centre of Excellence for Gravitational Wave Discovery (OzGrav)}
\affiliation{Gravitational Wave Data Centre, Swinburne University, Hawthorn VIC 3122, Australia}

\author{Marco Drago}
\affiliation{Universit\`a di Roma ``La Sapienza,'' I-00185 Roma, Italy}
\affiliation{INFN, Sezione di Roma, I-00185 Roma}

\author{Patrick Godwin}
\affiliation{Department of Physics, The Pennsylvania State University, University Park, PA 16802, USA}
\affiliation{Institute for Gravitation and the Cosmos, The Pennsylvania State University, University Park, PA 16802, USA}

\author{Shaon Ghosh}
\affiliation{Department of Physics and Astronomy, Montclair State University, Montclair, NJ, 07043}

\author{Giuseppe Greco}
\affiliation{Universit\`a degli Studi di Urbino "Carlo Bo", I-61029 Urbino, Italy  }
\affiliation{INFN, Sezione di Firenze, I-50019 Sesto Fiorentino, Firenze, Italy}
\author{Chad Hanna}
\affiliation{Department of Physics, The Pennsylvania State University, University Park, PA 16802, USA}
\affiliation{Institute for Gravitation and the Cosmos, The Pennsylvania State University, University Park, PA 16802, USA}
\affiliation{Department of Astronomy and Astrophysics, The Pennsylvania State University, University Park, PA 16802, USA}

\author{Shasvath J. Kapadia}
\affiliation{International Centre for Theoretical Sciences, Tata Institute of Fundamental Research, Bangalore 560089, India}

\author{Erik Katsavounidis}
\affiliation{LIGO Laboratory, Massachusetts Institute of Technology, Cambridge, MA 02139, USA}
\affiliation{Department of Physics and Kavli Institute for Astrophysics and Space Research,
Massachusetts Institute of Technology, Cambridge, MA 02139, USA}

\author{Victor Oloworaran}
\affiliation{Australian Research Council Centre of Excellence for Gravitational Wave Discovery (OzGrav)}
\affiliation{Department of Physics, University of Western Australia, Crawley WA 6009, Australia}

\author{Alexander E. Pace}
\affiliation{Department of Physics, The Pennsylvania State University, University Park, PA 16802, USA}
\affiliation{Institute for Gravitation and the Cosmos, The Pennsylvania State University, University Park, PA 16802, USA}

\author{Fiona Panther}
\affiliation{Australian Research Council Centre of Excellence for Gravitational Wave Discovery (OzGrav)}
\affiliation{Department of Physics, University of Western Australia, Crawley WA 6009, Australia}

\author{Anwarul Patwary}
\affiliation{Australian Research Council Centre of Excellence for Gravitational Wave Discovery (OzGrav)}
\affiliation{Department of Physics, University of Western Australia, Crawley WA 6009, Australia}

\author{Roberto De Pietri}
\affiliation{Dipartimento di Scienze Matematiche, Fisiche e Informatiche, Universit\`a di Parma, I-43124 Parma, Italy  }
\affiliation{7INFN, Sezione di Milano Bicocca, Gruppo Collegato di Parma, I-43124 Parma, Italy}

\author{Brandon Piotrzkowski}
\affiliation{Leonard E. Parker Center for Gravitation, Cosmology, and Astrophysics, University of Wisconsin-Milwaukee, Milwaukee, WI 53201, USA}

\author{Tanner Prestegard}
\affiliation{Leonard E. Parker Center for Gravitation, Cosmology, and Astrophysics, University of Wisconsin-Milwaukee, Milwaukee, WI 53201, USA}

\author{Luca Rei}
\affiliation{INFN, Sezione di Genova, I-16146 Genova, Italy  }

\author{Anala K. Sreekumar}
\affiliation{Australian Research Council Centre of Excellence for Gravitational Wave Discovery (OzGrav)}
\affiliation{Department of Physics, University of Western Australia, Crawley WA 6009, Australia}

\author{Marek~J.~Szczepa\'nczyk}
\affiliation{University of Florida, Gainesville, FL 32611, USA}

\author{Vinaya Valsan}
\affiliation{Leonard E. Parker Center for Gravitation, Cosmology, and Astrophysics, University of Wisconsin-Milwaukee, Milwaukee, WI 53201, USA}

\author{Aaron Viets}
\affiliation{Leonard E. Parker Center for Gravitation, Cosmology, and Astrophysics, University of Wisconsin-Milwaukee, Milwaukee, WI 53201, USA}

\author{Madeline Wade}
\affiliation{Department of Physics, Kenyon College, Gambier, OH 43022, USA}

\author{Linqing Wen}
\affiliation{Australian Research Council Centre of Excellence for Gravitational Wave Discovery (OzGrav)}
\affiliation{Department of Physics, University of Western Australia, Crawley WA 6009, Australia}

\author{John Zweizig}
\affiliation{LIGO, California Institute of Technology, Pasadena, CA 91125, USA}

\author{Other opt-ins}

\begin{abstract}

Gravitational-wave observations became commonplace in Advanced
LIGO-Virgo's recently concluded third observing run. {\NumPublicAlerts}
non-retracted candidates were identified and publicly announced in near real time. Gravitational waves from binary
neutron star mergers, however, remain of special interest since they can be
precursors to high-energy astrophysical phenomena like $\gamma$-ray bursts and
kilonovae. While late-time electromagnetic emissions provide important
information about the astrophysical processes within, the prompt emission along with
gravitational waves uniquely reveals the extreme matter and gravity during - and in the seconds following - merger. Rapid communication of source location and properties
from the gravitational-wave data is crucial to facilitate multi-messenger
follow-up of such sources. This is especially enabled if the partner facilities
are forewarned via an \emph{early-warning} (pre-merger) alert. Here we describe
the commissioning and performance of such a low-latency infrastructure within
LIGO-Virgo. We present results from an end-to-end mock data challenge that
detects binary neutron star mergers and alerts partner facilities before
merger. We set expectations for these alerts in future observing runs.
\end{abstract}


\section{Introduction}

The field of gravitational-wave astronomy has exploded in the years following
the first direct observation of \acp{GW} from a \ac{BBH}
merger~\citep{Abbott:2016blz}. Since then, LIGO-Virgo have published
\NumPublishedCandidates{} candidate events, many of which were identified in
low-latency\footnote{Some of the {\NumPublicAlerts} have not yet appeared in a
LIGO-Virgo publication.}; these include \NumBNS{} \ac{BNS} and \NumNSBH{}
\ac{NSBH} candidates~\citep{Abbott:2020niy}. The detection of
\acp{GW} from compact binaries, especially from \acp{BBH}, has become routine.
\ac{GW}s from \ac{BNS} and \ac{NSBH} mergers, however, remain rare. \ac{BNS}
and \ac{NSBH} mergers are of special interest due to the
possibility of counterpart \ac{EM} signals. For \ac{BNS} mergers in particular, it has long
been hypothesized that the central engine (post-merger) can launch
\acp{SGRB}~\citep{Lattimer:1976kbf, Lee:2007js}, kilonovae \citep{Li:1998bw,
Metzger:2010sy}, and radio waves and X-rays post merger~\citep{Nakar:2011cw,Metzger:2011bv}.
In the special case of the presence of a magnetized NS, it can also lead to GRB
precursors before the merger~\citep{Metzger:2016mqu}.

Although the improvement in Advanced LIGO-Virgo's sensitivity was paralleled by
analogous advancements in the field of time-domain astronomy, the first
observed \ac{BNS} merger, GW170817~\citep{TheLIGOScientific:2017qsa}, remains
the only realization of \ac{MMA} with \acp{GW}. The coincident observation of
\acp{GW} followed by an \ac{SGRB}, GRB~170817A, and the kilonova
AT~2017gfo,~\citep{GBM:2017lvd} bore evidence to the several-decade-old
hypothesis that compact object mergers were progenitors of these exotic
transients. The joint observations also contributed greatly to our
understanding of fundamental physics~\citep{Monitor:2017mdv,
LIGOScientific:2019fpa} and astrophysical processes associated with extreme
environments~\citep{Abbott:2017wuw, Nicholl:2017ahq}. Despite the plethora of
late-time observations made starting $\sim8$ hours after
coalescence~\citep{GBM:2017lvd}, observations of the prompt spectra were
precluded by non-stationarities in the LIGO Livingston interferometer and
delays in Virgo data transfer. The alert and sky localization were distributed
to partner observatories $\sim 40$ minutes~\citep{GCN21505} and $\sim 5$
hours~\citep{GCN21513}, respectively, after the signal arrived at the
detectors; by this time, the source had set below the horizon for northern
hemisphere telescopes. The circumstances surrounding this delay were
unusual, but it is crucial for LIGO-Virgo
to distribute alerts as quickly as possible to maximize the chance of
additional multi-messenger observations.

The serendipitous discovery of GRB~170717A by Fermi and INTEGRAL show the
importance of catching the prompt EM emission to our understanding of merging
compact binaries. EM observatories have begun to develop capacity to perform
targeted observations in response to preliminary \ac{GCN} notices produced by
pre-merger detections. For example, the
Murchison Wide-Field Array (MWA) radio telescope has a large field of view
ideally suited to searching for precursor and prompt radio emission from GW
sources and an established observing plan to respond to pre-merger
detections~\citep{James:2019xca}. \textit{Swift}-BAT has recently also
demonstrated the potential to respond autonomously to extremely low-latency
triggers in the future, with the introduction of an on-board sub-threshold
trigger recovery algorithm (GUANO,~\cite{Tohuvavohu:2020stm}). By the beginning
of Advanced LIGO-Virgo's fourth observing run (O4), it is expected that established missions and observatories will be
joined by next generation facilities like the Rubin
Observatory~\citep{Ivezic:2008fe}. This greatly improves the chances of
performing targeted followup observations of prompt, or even
precursor~\citep{troja2010precursors,tsang2012resonant}, emission from compact
binary mergers provided that pre-merger alerts can be
issued.

LIGO-Virgo has since streamlined the alert process (see
Fig.~\ref{fig:end_to_end_latencies}). \ac{O3} saw the dawn of autonomously
distributed Preliminary \ac{GCN} Notices~\citep{GCN24045}\footnote{
\url{https://gcn.gsfc.nasa.gov/}}, which allowed LIGO-Virgo to notify the world
of candidate signals within $7.0^{+92}_{-4}$ minutes\footnote{The 95\%
reported here is severely impacted by several high latency events that evaded
automated procedures.} of
observation. To further enable EM-GW observations, we can leverage the long-lived nature of
\acp{BNS} in the sensitive band of advanced ground-based \ac{GW}
detectors to make pre-merger detections~\citep{Cannon:2011vi,
10.1093/mnras/stw576}. This was recently demonstrated by
\citet{Sachdev:2020lfd} and \citet{Nitz:2020vym}. The early detection and communication of \acp{GW}
from \acp{BNS} aims to facilitate \ac{EM} follow-up efforts by further
reducing the latency of alerts and improving prospects of capturing the initial
spectra.

In this letter we describe the commissioning and performance of the
low-latency sub-system within Advanced LIGO-Virgo that is able to provide
pre-merger alerts for electromagnetically bright compact binaries. We begin by
describing the end-to-end low-latency workflow in Section~\ref{sec:analysis}, from the
time of data acquisition to the dissemination of public alerts. We then assess the
performance of a subset of this infrastructure in a mock data challenge
described in Section~\ref{sec:results}, with special emphasis placed on
pre-merger alerts. We demonstrate that Preliminary \ac{GCN} Notices can be
distributed with true negative latencies: partner observatories receive sky
localizations and source information before the binary has completed its
merger. We report on the improved latencies at each step of the workflow, and
set expectations for pre-merger alerts in O4 and next generation detectors in Section~\ref{sec:prospects}.

\section{Analysis}\label{sec:analysis}

The low-latency workflow begins with data acquisition at each interferometer.
The digital signal from the output photodiode is initially calibrated by a
pipeline that runs on the set of computers that directly control the
interferometer. The calibrated data, while produced with near-zero latency,
are not yet accurate enough for use by low-latency gravitational-wave searches.
The data are broadcast to a set of computers
where a GStreamer-based pipeline corrects the
strain data to achieve the required level of accuracy~\citep{Viets2018}. This pipeline writes
the calibrated strain data to a proprietary LIGO frame data format and then
transfers them to computing sites.
There, the calibrated data are ingested by the complete set of low-latency full bandwidth
\ac{GW} pipelines:
cWB~\citep{Klimenko:2004qh,Klimenko:2005xv,Klimenko:2006rh,Klimenko:2011hz,Klimenko:2015ypf},
GstLAL~\citep{Sachdev:2019vvd, Hanna:2019ezx, Messick:2016aqy},
MBTAOnline~\citep{mbta}, PyCBC Live~\citep{PyCBCLiveO2,PyCBCLiveO3},
and SPIIR~\citep{Luan2012, Hooper2012, Liu2012, Guo2018, spiir}. For the
first time, we also incorporate two matched-filter based pipelines focused on
pre-merger detection into our workflow:
GstLAL~\citep{Sachdev:2020lfd,Cannon:2011vi} and SPIIR~\citep{Chu:2020pjv}. All detection
pipelines analyze the data for \ac{GW}s and assign significances to
candidate triggers.
Candidates that are assigned \acp{FAR} less
than one per hour\footnote{No trials factor is applied to the candidate upload
threshold.} are uploaded to the \ac{GraceDB}~\footnote{\url{https://gracedb.ligo.org/}}
alongside data required downstream in the alert process.

After candidates are uploaded, the task manager
{\gwcelery}~\footnote{\url{https://gwcelery.readthedocs.io/}} interacts with
low-latency searches and {\gracedb} to orchestrate a number of parallel and
interconnected processes which, in the event of a discovery, culminates in the
dissemination of GCN Notices. {\gwcelery} provided the semi-automated
infrastructure for public alerts in O3, as well as for the mock data challenge
reported here. The major subsystems include:
\begin{itemize}
	\item The listener for {\lvalert}, which is a publish-subscribe system used 
          by {\gracedb} to push machine-readable notifications about its state.
	\item The Superevent Manager, which clusters and merges related candidates into
	      \emph{superevents}.\footnote{\url{https://emfollow.docs.ligo.org/userguide/analysis/superevents.html}}
	\item The client functionality to interact with {\gracedb}.\footnote{
	      \url{https://gracedb-sdk.readthedocs.io}}
	\item The GCN listener that listens for notices from external facilities
	      to spot coincidences with \ac{GW} candidates.
	\item The External Trigger Manager, which correlates gravitational-wave events
	      with GRB, neutrino, and supernova events.
	\item The GCN broker that disseminates \ac{GW} candidate information for external
	      consumption.
	\item The Orchestrator, which executes the per-(super)event annotation workflow.
\end{itemize}

\begin{figure*}
    \centering
	\includegraphics[width=1.1\textwidth, trim=1cm 0cm 0cm 0cm]{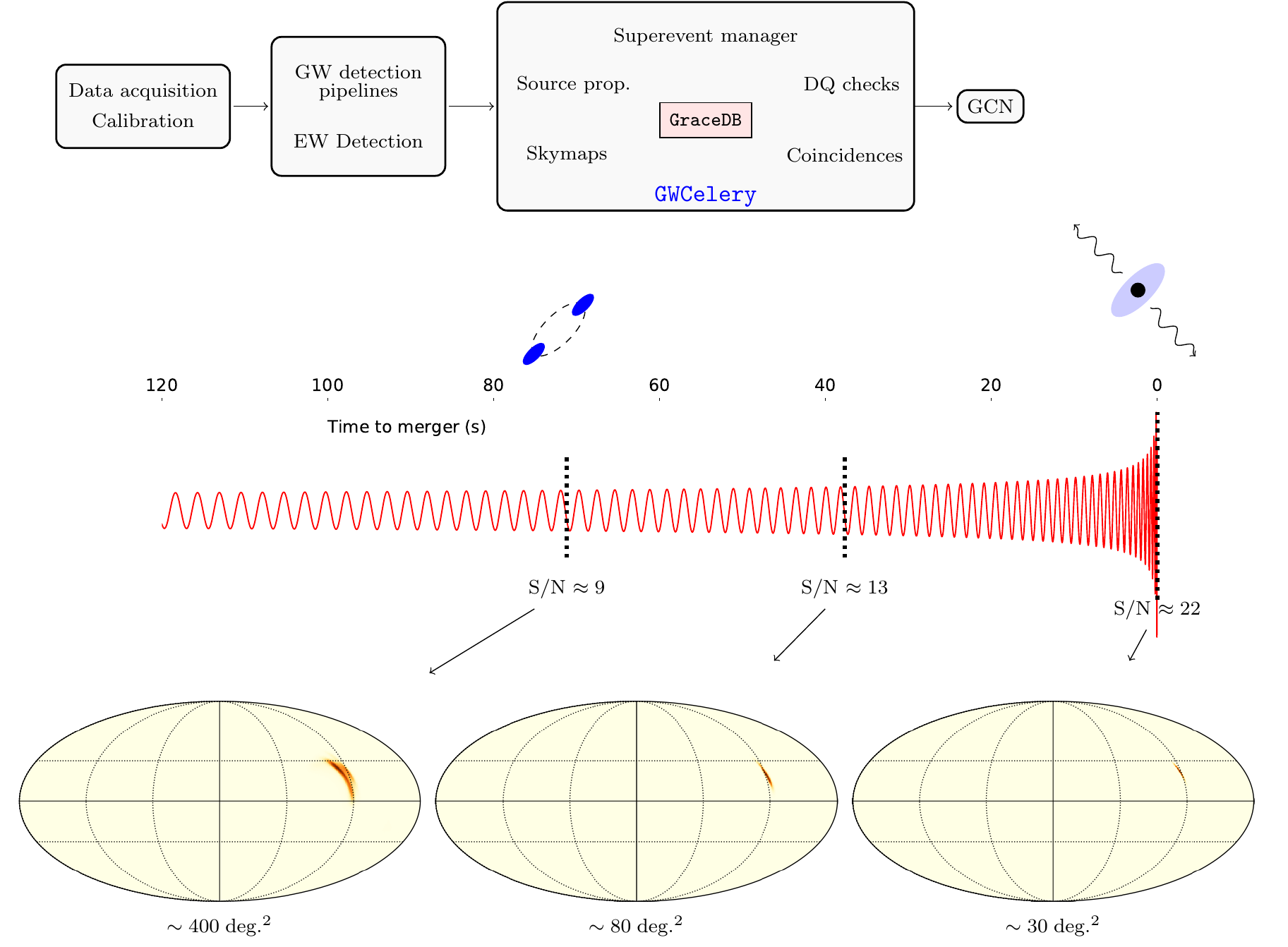}
    \caption{\label{fig:interactions}
	This upper half of the figure illustrates the complete pipeline and interaction of
	the various (sub)systems, mentioned in Sec.~\ref{sec:analysis}, responsible for
	disseminating early warning alerts. The waveform evolution with time is shown in
	the bottom half along with the dependence of the sky-localization area on the cutoff
	time of the early-warning templates and the accumulated \ac{SNR}
	during the binary inspiral. The waveforms, time to merger, \ac{SNR},
	and localizations in this figure are qualitative.
    }
\end{figure*}

After candidate events are uploaded by detection pipelines, they are
localized via BAYESTAR~\citep{PhysRevD.93.024013}, given a probability
of having an electromagnetic counterpart~\citep{Chatterjee:2019avs}, and
assigned a source-category based astrophysical probability under the
assumption that astrophysical and terrestrial triggers
occur as independent Poisson processes~\citep{Kapadia:2019uut}. Events are
checked for temporal and, when possible, spatial coincidences with gamma-ray
bursts or neutrino bursts using the RAVEN pipeline~\citep{2016PhDT.........8U}.
A joint significance is calculated to decide whether the joint candidate should
be published.

BAYESTAR was optimized in order to support early warning localizations which led to
a median run time of 0.5\,s per event for early warning triggers and 1.1\,s per event
for full bandwidth triggers. The latter is a $4.2\times$ speedup compared to
usual O3 performance. The significant changes included rearrangement of loops
to improve memory access patterns and make better use of x86\_64 vector instructions,
changes to the input data handling to distinguish properly between the merger
time and the cutoff time of early warning templates, and the redesign of the reconstruction
filter that is used to sample the SNR time series for likelihood evaluation to use
a lower sample rate.~\footnote{The early warning templates are Nyquist critically sampled which
could lead to ringing artifacts.}

To mitigate the effect of noise transients, basic data quality checks are also
performed for every candidate uploaded to {\gracedb}. In
particular, specific state vectors are checked to ensure that candidate events
occur during times when the relevant
detectors are in observing mode and to verify that there are no
coincident hardware injections.

A qualitative overview of entire pipeline and the various (sub)systems mentioned
above is illustrated in Fig.~\ref{fig:interactions}. A heuristic waveform evolution
and the effect of different early-warning template cutoff times on the
accumulated \ac{SNR} and the sky-localization is also shown.

\section{Results}\label{sec:results}

\begin{figure*}[ht!]
	\includegraphics[width=\textwidth]{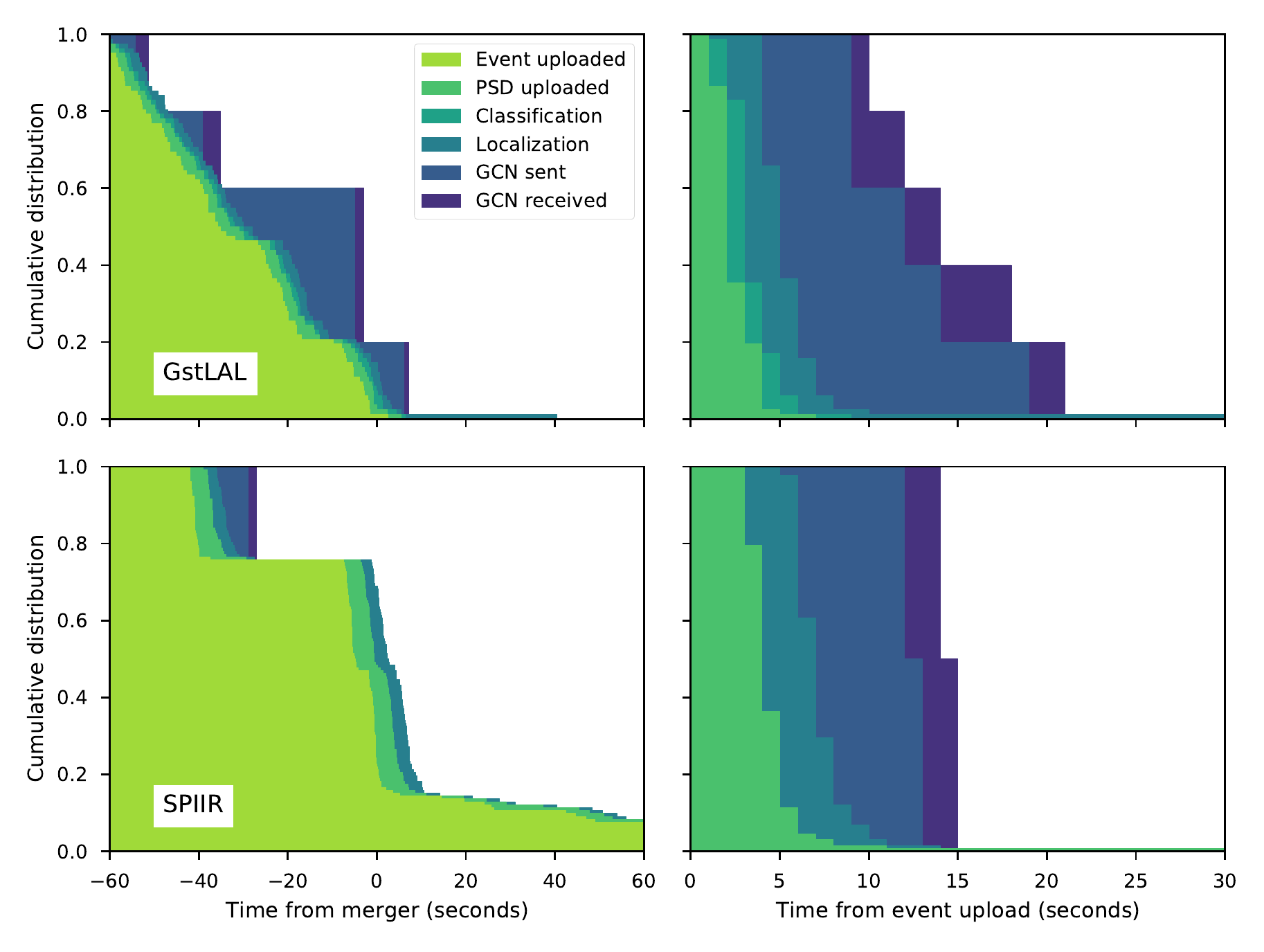}
	\caption{\label{fig:latencies}
	Latencies associated with early warning uploads from the GstLAL (top) and SPIIR
	(bottom) pipelines. Design differences between the pipelines lead to distinct
	distributions for the time before merger at which a candidate is identified.
	The left panels indicate that $\sim 85\%$ and $\sim 35\%$ of the uploaded GstLAL
	and SPIIR candidates, respectively, are localized prior to merger. The
	right panels demonstrate that despite differences in latencies associated with
	event identification, the scatter of the remaining processes is remarkably
	similar.}
\end{figure*}

To demonstrate the robustness of the alert infrastructure, we describe the
results of a mock data challenge carried out between 11 June 2020 1700 UTC and
19 June 2020 1700 UTC. Data previously collected during O3 were replayed
as a mock low-latency analysis. We note that since
the challenge relied on previously collected data, it was impossible to test the full
low-latency workflow; notably, data transfer and calibration latencies are not
included ($\sim 5$ seconds). The test therefore begins with the detection pipelines, but
otherwise follows a workflow identical to Advanced LIGO-Virgo observing runs.

The \ac{FAR} threshold set for issuing
early warning test notices was chosen to be \EWFARThresh{}. Full bandwidth
triggers used the same FAR threshold set throughout O3 for public alerts
(\FARThresh{})\footnote{A trials factor is applied on top of this
threshold to account for the two early warning and four full bandwidth matched
filter pipelines}.
At fixed \ac{FAR}, the astrophysical probability~\citep{Kapadia:2019uut}
associated with pre-merger analyses is lower than for full bandwidth analyses.
Due to this fact, combined with our chosen higher \ac{FAR} threshold for
early-warning alerts, we issued retraction circulars for early warning
candidates that were not also identified by the full bandwidth analyses. There
were no retraction criteria set for full bandwidth triggers.

During the mock data challenge, eight candidates were published
via the test \ac{GCN}. \NumFullBandCircs{} candidates were identified by
only the full bandwidth analyses and were distributed via notice and 
circular~\citep{GCN27977,GCN27965,GCN27963}. The remaining \NumEWRetractions{}
public candidates were identified only by the early warning pipelines and were
distributed via \ac{GCN} notices to subscribers of test alerts. None of these
\NumEWRetractions{} candidates were observed in the full bandwidth analyses;
they were therefore subsequently
retracted~\citep{GCN27951,GCN27987,GCN27988,GCN27989,GCN27990}.
Out of the \NumEWRetractions{} retracted triggers, \NumGstLALEWRetractions{} came
from the GstLAL early warning pipeline, while \NumSPIIREWRetractions{}
was issued by the SPIIR early warning pipeline. An
authentication issue prevented the SPIIR pipeline from issuing additional
events past the \ac{FAR} threshold. A summary of the \NumEWRetractions{} early warning alerts is
given in Table~\ref{tab:summary}.

Although only \NumEWRetractions{} pre-merger candidates passed the early
warning public alert threshold, GstLAL and SPIIR uploaded \NumGstLALEWUploads{}
and \NumSPIIREWUploads{} early warning candidate events, respectively, to \ac{GraceDB}. We use the
metadata associated with these uploads to produce Fig.~\ref{fig:latencies}. 
From the events crossing threshold we see that the maximum delivery time from event upload is 15s, independent of pipeline.
This enables $\sim 85\%$ and $\sim 35\%$ of the GstLAL and SPIIR
candidates, respectively, to be localized before merger.

\begin{figure*}
	\includegraphics[width=\textwidth]{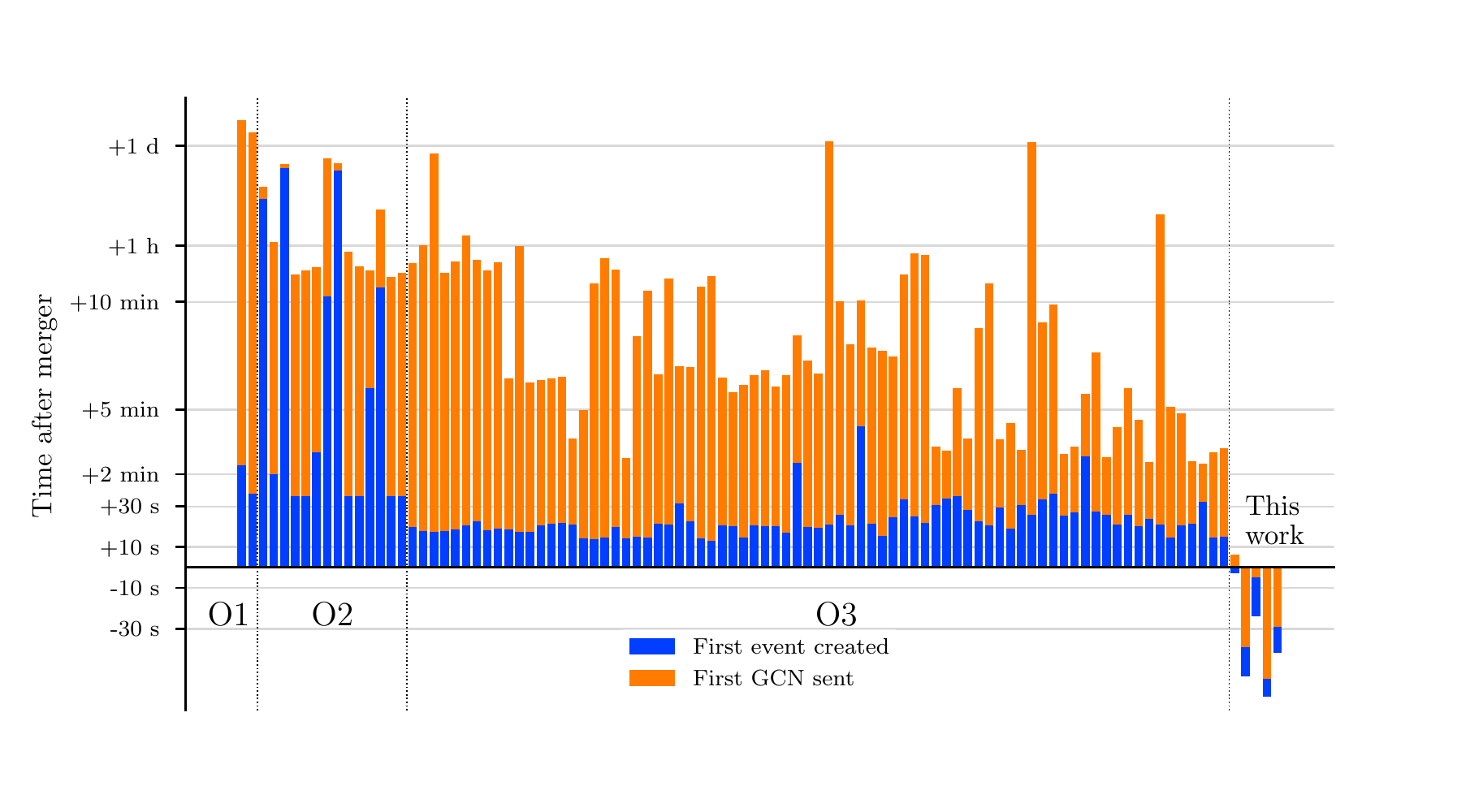}
    \caption{\label{fig:end_to_end_latencies} A history of end-to-end latencies across public alerts in the first three observing runs and the mock data challenge presented here \citep{LIGOScientific:2019gag}.}
\end{figure*}

\section{Looking ahead}\label{sec:prospects}

Early warning alerts using real data have not yet been released by the LIGO-Virgo
collaboration.  Despite the steady improvement of the alert infrastructure
(Figure~\ref{fig:end_to_end_latencies}), there remain several areas for
improvement in the processing of data and production of alerts if the
collaboration decides to pursue pre-merger triggers.
As previously mentioned, low-latency data calibration is currently a two step
process; the near-zero-latency pipeline is corrected by a secondary
GStreamer-based pipeline. Work is underway to reduce this to a single
calibration step to reduce latency by $\mathcal{O}$(seconds). The calibrated data are transferred from
the detector sites to the computing clusters in $\sim 4$ seconds, and afterward at the
cluster level using Kafka,\footnote{\url{https://kafka.apache.org/}} with an additional
$\sim 0.1$ seconds. Another one second of latency~\footnote{Four seconds for Virgo data.} is attributed to
the choice to distribute data via frame files.
A number of improvements are under development to reduce this latency budget.

Reductions to the noise budget at frequencies $\lesssim 30\thinspace\mathrm{Hz}$ will
improve the possibility of detection pipelines identifying signals long before
merger. We estimate that if the noise floor below $30\thinspace\mathrm{Hz}$ remains
unchanged from O3, the recovered \ac{SNR} one minute and 30 seconds before
merger will be $\sim50\%$ and $\sim20\%$ less, respectively, than if the detectors
reach the previously projected O4 sensitivity. The effect
is less severe for early warning times just before merger, but low frequency
noise is a major barrier to advance alerts.

\begin{figure*}[ht!]
\centering
	\subfloat[]{%
		\includegraphics{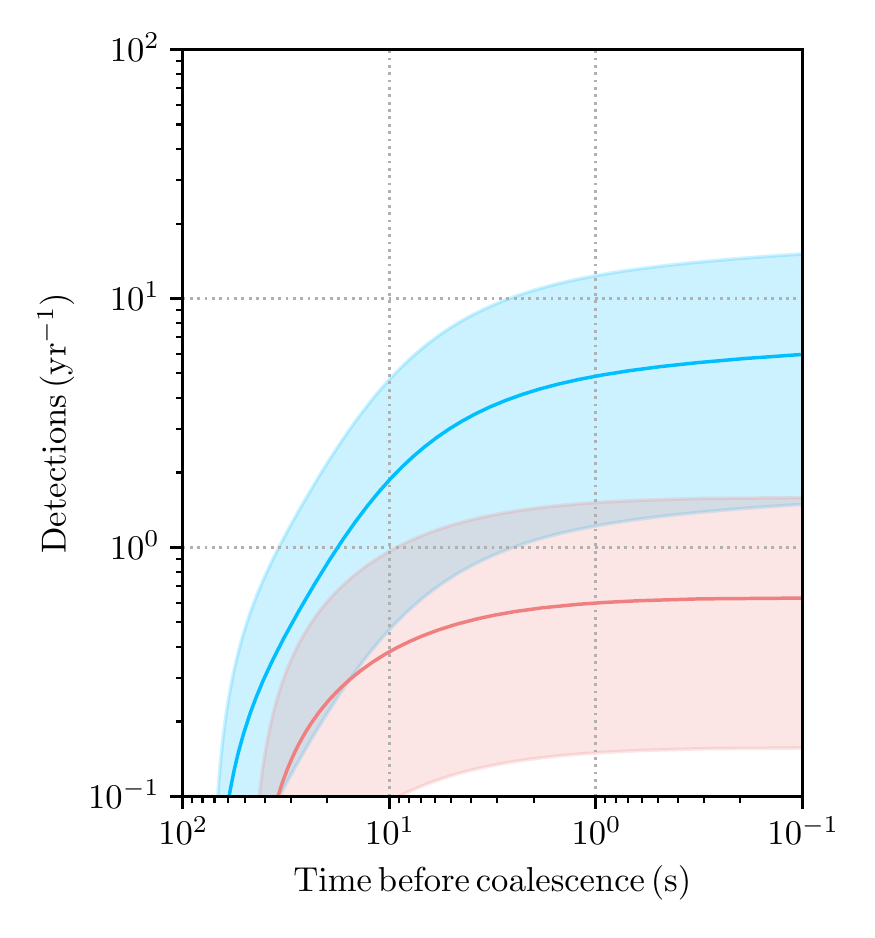}
			\label{fig:o4detections}
}
	\subfloat[]{%
		\includegraphics{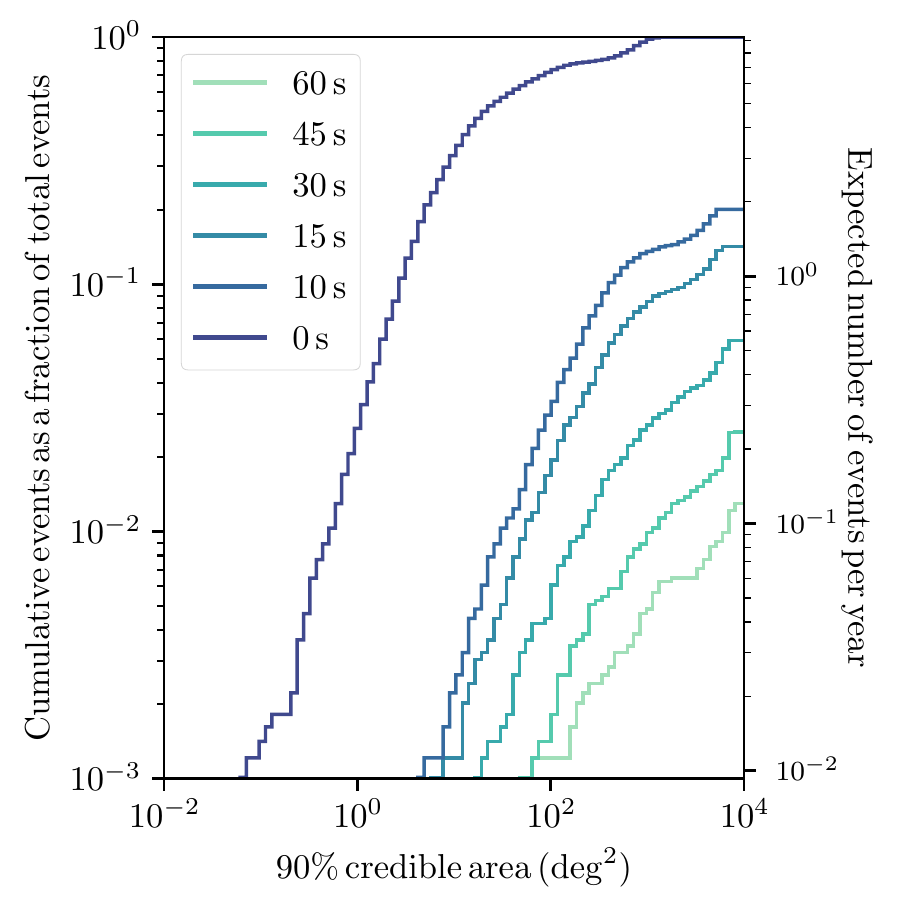}
			\label{fig:o4localizations}
}
	\caption{(\ref{fig:o4detections}) Projected O4 early warning detection
rate assuming 0 second (blue) and 25 second (red) end-to-end latencies from the
GW alert system. The worst case scenario assumes 5 seconds for calibration and
data transfer, 5 seconds for pipeline analysis, and 15 seconds for event upload
and GCN creation. The rate of expected detections was estimated from a
simulated data set assuming a 100\% detector duty cycle for the 4-detector HLVK
network. The uncertainty bands reflect the $(5\%,95\%)$ confidence region for
the \ac{BNS} rate. Signals with network \ac{SNR}s greater than 12 are considered
recovered.  (\ref{fig:o4localizations}) The expected localization distribution
for \ac{BNS} detections at six approximate early warning times. No latencies
are included in this figure. The inclusion of an end-to-end latency does not
shift the histogram itself; the labeled times before merger would all
systematically shift instead. Both plots use the \ac{BNS} rates estimated
in~\cite{Abbott:2020gyp}.}
\end{figure*}

Figures~\ref{fig:latencies} and ~\ref{fig:end_to_end_latencies} demonstrate
that the \ac{GW} alert system is capable of providing \ac{GW} alerts before
merger, but they do not consider the prospects for detection from an
astrophysical source population. We generate a population of simulated \ac{BNS}
signals, henceforth referred to as \textit{injections}, using the
\texttt{TaylorF2}~\citep{Sathyaprakash:1991mt, Blanchet:1995ez,
Blanchet:2005tk, Buonanno:2009zt} waveform model. Both source-frame component
masses are drawn from a Gaussian distribution between \SimMinMass{} $< m_1, m_2
<$ \SimMaxMass{} with mean mass of \SimMeanMass{} and standard deviation of
\SimStdev{}, modeled after observations of galactic
\acp{BNS}~\citep{Ozel:2016oaf}\footnote{Note that if GW190425 is a \ac{BNS}, then
galactic measurements are not representative of neutron star masses.}.
The neutron stars in the population are non-spinning, motivated by the low
spins of \acp{BNS} expected to merge within a Hubble
time~\citep{Burgay:2003jj,Zhu:2017znf}. The signals are distributed uniformly
in comoving volume up to a redshift of $z = \SimMaxRedshift{}$. We consider a
network of four GW detectors: LIGO-Hanford, LIGO-Livingston,
Virgo, and KAGRA at their projected O4 sensitivities.\footnote{
\url{https://dcc.ligo.org/LIGO-T2000012/public}} We simulate the results of
an early warning matched-filtering pipeline by considering 6 different discrete
frequency cut-offs: $29\thinspace\rm{Hz}$, $32\thinspace\rm{Hz}$,
$38\thinspace\rm{Hz}$, $49\thinspace\rm{Hz}$, $56\thinspace\rm{Hz}$, and
$1024\thinspace\rm{Hz}$ to analyze signal recovery at (approximately)
$58\thinspace\rm{s}$, $44\thinspace\rm{s}$, $28\thinspace\rm{s}$,
$14\thinspace\rm{s}$, $10\thinspace\rm{s}$, and $0\thinspace\rm{s}$ before
merger, motivated by~\cite{Sachdev:2020lfd}. We calculate the network \ac{SNR} of
each injection at each frequency cut-off and consider the events that pass an
\ac{SNR} cut-off of 12.0
as `detected'.  We then calculate the sky posteriors for each of the detected
signals by using BAYESTAR~\citep{PhysRevD.93.024013}. We use the most
recent \ac{BNS} local merger rate from~\cite{Abbott:2020gyp} of $320^{+410}_{-240}
\, \text{Gpc}^{-3}\text{yr}^{-1}$ to estimate the number of events detected per
year in the detector network. In Figure~\ref{fig:o4detections} we see that our
optimistic scenario predicts $5^{+7}_{-4}$ \ac{GCN} will be received $1$ second
before merger per year, while our pessimistic scenario
predicts $\mathcal{O}(1)$ \ac{GCN} will be received $1$ second before merger
per year considering the higher end of the \ac{BNS} rate.
Figure~\ref{fig:o4localizations} predicts that $\sim 9$ events will be detected
per year, out of which $\sim 20\%$ ($\sim 1.3\%$) will be detected $10\, \rm s$
$(60 \, \rm s)$ before merger. Further, $\sim 3\%$ of the detectable events
($\sim1$ \ac{BNS} every 3--4 years) will be detected 10 seconds prior to merger
and have a localization less than 100
$\mathrm{deg}^2$ at O4 sensitivities. This
highlights the need for continued latency improvements in advance of O4 to
maximize the potential of capturing prompt emission.

In the design sensitivity era with three detectors, \cite{Sachdev:2020lfd} have
shown that about half of the total detectable \acp{BNS} will be found $10\rm \,
s$ before merger, and about 2\% will be identified before merger and localized
to within 100 $\text{deg}^2$. \cite{Sachdev:2020lfd} used the GstLAL pipeline
in an early warning configuration to assign \ac{FAR}s to simulated \ac{BNS}
signals to estimate these rates.\footnote{Note that the estimated \ac{BNS} rate
at the time of~\cite{Sachdev:2020lfd} was
approximately three times larger than the updated rate presented
in~\cite{Abbott:2020gyp}} We extend this to include KAGRA in the detector
network, but we
estimate rates based on a fiducial \ac{SNR} cut-off of 12. We find that our zero-latency
scenario improves to $\sim2$ \ac{BNS} observable one minute before coalescence.
Assuming 25 seconds of pipeline latency, $\sim 1$ \ac{BNS} will be localized and
disseminated one minute before merger every 2 years. The localization prospects similarly
improve. At design sensitivity, $\sim 3$ \ac{BNS} every year will be detected 10 seconds prior
to merger and have localizations $\lesssim 100 \text{deg}^2$, $\sim 2$ signals per
year will be detected 15 seconds prior to merger with similar localization.
The detection rates estimated by \cite{Nitz:2020vym} are comparable to ours,
considering their use of a larger \ac{BNS} rate density ($\sim 3$ times ours)
and a less strict criterion for the detectability of a signal (network \ac{SNR} $> 10$).

The next generation of ground based interferometers will offer unparalleled
early warning capabilities. Using a similar \ac{SNR} detection threshold (but
further mandating that at least two interferometers measure \ac{SNR}s above
5.5), ~\citet{Chan:2018csa} found that the Einstein Telescope can alert
observers up to 20 hours in advance for 58\% of detectable \ac{BNS} at 200\mpc{} and
100\% at 40\mpc{}. The majority of these signals will be well localized. A
similar study by~\citet{Akcay:2018aqh} with a \ac{SNR} detection threshold of 15 found that the
Einstein Telescope will provide early notice for $\mathcal{O}(10^2)$ \ac{BNS}
mergers next decade. 

\acknowledgments

We are grateful to B.S. Sathyaprakash for reviewing our manuscript and
providing useful comments. We thank the LIGO Laboratory for use of its
computing facility to make this work possible, and we gratefully acknowledge
the support of NSF grants PHY-0757058 and PHY-0823459. C.H. gratefully
acknowledges the support of NSF grant OAC-1841480.  DC acknowledges NSF grant
no. PHY-1700765 and PHY-1912649, and is supported by the Illinois Survey
Science Fellowship of the Center for Astrophysical Surveys (CAPS) at the University of
Illinois Urbana-Champaign. S.S. is supported by the Eberly Research Funds of
Penn State, The Pennsylvania State University, University Park, Pennsylvania.
G.~M. is supported by the National Science Foundation (NSF) through award
PHY-1764464 to the LIGO Laboratory. MK, QC, FP, LW, AP, AS, VO acknowledge the
funding from Australian Research Council (ARC) Centre of Excellence for
Gravitational Wave Discovery OzGrav under grant CE170100004. 

\facilities{LIGO, EGO:Virgo}

\clearpage
\software{astropy \citep{2013A&A...558A..33A}, numpy \citep{2020NumPy-Array},
    matplotlib \citep{matplotlib}, iPython \citep{ipython},
    pandas \citep{pandas}, gwpy \citep{gwpy}, celery \citep{celery}
}
\input{acronyms}

\FloatBarrier
\appendix
\begin{deluxetable*}{ccc|cccc|c}[h]
    \label{tab:summary}
    \tabletypesize{\scriptsize}
    \centerwidetable
    \tablewidth{1.0\columnwidth}
	\tablecaption{A summary of the 5 early warning alert information and latencies
	from the mock data challenge described in Sec.~\ref{sec:results}.
	Among the 5, MS200619bf was reported by the SPIIR pipeline, while
	the others were reported from GstLAL. The latencies are broken down
	in steps of the event being uploaded into GraceDB, the superevent
	being created, the skymap being available for the preferred event,
	and the notice being acknowledged by GCN.}
	\tablehead{
		\colhead{Superevent} &
		\colhead{Date (UTC)} &
		\colhead{FAR} &
		\multicolumn{4}{c}{Latency} &
		\colhead{GCNs} \\
		& & & Event & Superevent & Skymap & Notice &
	}
	\startdata
	\input{table_1_data}
	\enddata
\end{deluxetable*}

\bibliography{early-warning-paper}{}
\bibliographystyle{aasjournal}

\end{document}

%% file: acronyms.tex
\providecommand{\acrolowercase}[1]{\lowercase{#1}}

\begin{acronym}
\acro{2D}[2D]{two\nobreakdashes-dimensional}
\acro{2+1D}[2+1D]{2+1\nobreakdashes--dimensional}
\acro{2MRS}[2MRS]{2MASS Redshift Survey}
\acro{3D}[3D]{three\nobreakdashes-dimensional}
\acro{2MASS}[2MASS]{Two Micron All Sky Survey}
\acro{AdVirgo}[AdVirgo]{Advanced Virgo}
\acro{AMI}[AMI]{Arcminute Microkelvin Imager}
\acro{AGN}[AGN]{active galactic nucleus}
\acroplural{AGN}[AGN\acrolowercase{s}]{active galactic nuclei}
\acro{aLIGO}[aLIGO]{Advanced \acs{LIGO}}
\acro{ASKAP}[ASKAP]{Australian \acl{SKA} Pathfinder}
\acro{ATCA}[ATCA]{Australia Telescope Compact Array}
\acro{ATLAS}[ATLAS]{Asteroid Terrestrial-impact Last Alert System}
\acro{BAT}[BAT]{Burst Alert Telescope\acroextra{ (instrument on \emph{Swift})}}
\acro{BATSE}[BATSE]{Burst and Transient Source Experiment\acroextra{ (instrument on \acs{CGRO})}}
\acro{BAYESTAR}[BAYESTAR]{BAYESian TriAngulation and Rapid localization}
\acro{BBH}[BBH]{binary black hole}
\acro{BHBH}[BHBH]{\acl{BH}\nobreakdashes--\acl{BH}}
\acro{BH}[BH]{black hole}
\acro{BNS}[BNS]{binary neutron star}
\acro{CARMA}[CARMA]{Combined Array for Research in Millimeter\nobreakdashes-wave Astronomy}
\acro{CASA}[CASA]{Common Astronomy Software Applications}
\acro{CBCG}[CBCG]{Compact Binary Coalescence Galaxy}
\acro{CFH12k}[CFH12k]{Canada--France--Hawaii $12\,288 \times 8\,192$ pixel CCD mosaic\acroextra{ (instrument formerly on the Canada--France--Hawaii Telescope, now on the \ac{P48})}}
\acro{CLU}[CLU]{Census of the Local Universe}
\acro{CRTS}[CRTS]{Catalina Real-time Transient Survey}
\acro{CTIO}[CTIO]{Cerro Tololo Inter-American Observatory}
\acro{CBC}[CBC]{compact binary coalescence}
\acro{CCD}[CCD]{charge coupled device}
\acro{CDF}[CDF]{cumulative distribution function}
\acro{CGRO}[CGRO]{Compton Gamma Ray Observatory}
\acro{CMB}[CMB]{cosmic microwave background}
\acro{CRLB}[CRLB]{Cram\'{e}r\nobreakdashes--Rao lower bound}
\acro{cWB}[\acrolowercase{c}WB]{Coherent WaveBurst}
\acro{DASWG}[DASWG]{Data Analysis Software Working Group}
\acro{DBSP}[DBSP]{Double Spectrograph\acroextra{ (instrument on \acs{P200})}}
\acro{DCT}[DCT]{Discovery Channel Telescope}
\acro{DECAM}[DECam]{Dark Energy Camera\acroextra{ (instrument on the Blanco 4\nobreakdashes-m telescope at \acs{CTIO})}}
\acro{DES}[DES]{Dark Energy Survey}
\acro{DFT}[DFT]{discrete Fourier transform}
\acro{EM}[EM]{electromagnetic}
\acro{ER8}[ER8]{eighth engineering run}
\acro{FD}[FD]{frequency domain}
\acro{FAR}[FAR]{false alarm rate}
\acro{FFT}[FFT]{fast Fourier transform}
\acro{FIR}[FIR]{finite impulse response}
\acro{FITS}[FITS]{Flexible Image Transport System}
\acro{F2}[F2]{FLAMINGOS\nobreakdashes-2}
\acro{FLOPS}[FLOPS]{floating point operations per second}
\acro{FOV}[FOV]{field of view}
\acroplural{FOV}[FOV\acrolowercase{s}]{fields of view}
\acro{FTN}[FTN]{Faulkes Telescope North}
\acro{FWHM}[FWHM]{full width at half-maximum}
\acro{GBM}[GBM]{Gamma-ray Burst Monitor\acroextra{ (instrument on \emph{Fermi})}}
\acro{GCN}[GCN]{Gamma-ray Coordinates Network}
\acro{GLADE}[GLADE]{Galaxy List for the Advanced Detector Era}
\acro{GMOS}[GMOS]{Gemini Multi-object Spectrograph\acroextra{ (instrument on the Gemini telescopes)}}
\acro{GRB}[GRB]{gamma-ray burst}
\acro{GraceDB}[GraceDB]{GRAvitational-wave Candidate Event DataBase}
\acro{GROWTH}[GROWTH]{Global Relay of Observatories Watching Transients Happen}
\acro{GSC}[GSC]{Gas Slit Camera}
\acro{GSL}[GSL]{GNU Scientific Library}
\acro{GTC}[GTC]{Gran Telescopio Canarias}
\acro{GW}[GW]{gravitational wave}
\acro{GWGC}[GWGC]{Gravitational Wave Galaxy Catalogue}
\acro{HAWC}[HAWC]{High\nobreakdashes-Altitude Water \v{C}erenkov Gamma\nobreakdashes-Ray Observatory}
\acro{HCT}[HCT]{Himalayan Chandra Telescope}
\acro{HEALPix}[HEALP\acrolowercase{ix}]{Hierarchical Equal Area isoLatitude Pixelization}
\acro{HEASARC}[HEASARC]{High Energy Astrophysics Science Archive Research Center}
\acro{HETE}[HETE]{High Energy Transient Explorer}
\acro{HFOSC}[HFOSC]{Himalaya Faint Object Spectrograph and Camera\acroextra{ (instrument on \acs{HCT})}}
\acro{HMXB}[HMXB]{high\nobreakdashes-mass X\nobreakdashes-ray binary}
\acroplural{HMXB}[HMXB\acrolowercase{s}]{high\nobreakdashes-mass X\nobreakdashes-ray binaries}
\acro{HSC}[HSC]{Hyper Suprime\nobreakdashes-Cam\acroextra{ (instrument on the 8.2\nobreakdashes-m Subaru telescope)}}
\acro{IACT}[IACT]{imaging atmospheric \v{C}erenkov telescope}
\acro{IIR}[IIR]{infinite impulse response}
\acro{IMACS}[IMACS]{Inamori-Magellan Areal Camera \& Spectrograph\acroextra{ (instrument on the Magellan Baade telescope)}}
\acro{IMR}[IMR]{inspiral-merger-ringdown}
\acro{IPAC}[IPAC]{Infrared Processing and Analysis Center}
\acro{IPN}[IPN]{InterPlanetary Network}
\acro{IPTF}[\acrolowercase{i}PTF]{intermediate \acl{PTF}}
\acro{IRAC}[IRAC]{Infrared Array Camera}
\acro{ISM}[ISM]{interstellar medium}
\acro{ISS}[ISS]{International Space Station}
\acro{KAGRA}[KAGRA]{KAmioka GRAvitational\nobreakdashes-wave observatory}
\acro{KDE}[KDE]{kernel density estimator}
\acro{KN}[KN]{kilonova}
\acroplural{KN}[KNe]{kilonovae}
\acro{LAT}[LAT]{Large Area Telescope}
\acro{LCOGT}[LCOGT]{Las Cumbres Observatory Global Telescope}
\acro{LHO}[LHO]{\ac{LIGO} Hanford Observatory}
\acro{LIB}[LIB]{LALInference Burst}
\acro{LIGO}[LIGO]{Laser Interferometer \acs{GW} Observatory}
\acro{llGRB}[\acrolowercase{ll}GRB]{low\nobreakdashes-luminosity \ac{GRB}}
\acro{LLOID}[LLOID]{Low Latency Online Inspiral Detection}
\acro{LLO}[LLO]{\ac{LIGO} Livingston Observatory}
\acro{LMI}[LMI]{Large Monolithic Imager\acroextra{ (instrument on \ac{DCT})}}
\acro{LOFAR}[LOFAR]{Low Frequency Array}
\acro{LOS}[LOS]{line of sight}
\acroplural{LOS}[LOSs]{lines of sight}
\acro{LMC}[LMC]{Large Magellanic Cloud}
\acro{LSB}[LSB]{long, soft burst}
\acro{LSC}[LSC]{\acs{LIGO} Scientific Collaboration}
\acro{LSO}[LSO]{last stable orbit}
\acro{LSST}[LSST]{Large Synoptic Survey Telescope}
\acro{LT}[LT]{Liverpool Telescope}
\acro{LTI}[LTI]{linear time invariant}
\acro{MAP}[MAP]{maximum a posteriori}
\acro{MBTA}[MBTA]{Multi-Band Template Analysis}
\acro{MCMC}[MCMC]{Markov chain Monte Carlo}
\acro{MLE}[MLE]{\ac{ML} estimator}
\acro{ML}[ML]{maximum likelihood}
\acro{MMA}[MMA]{multi-messenger astronomy}
\acro{MOU}[MOU]{memorandum of understanding}
\acroplural{MOU}[MOUs]{memoranda of understanding}
\acro{MWA}[MWA]{Murchison Widefield Array}
\acro{NED}[NED]{NASA/IPAC Extragalactic Database}
\acro{NIR}[NIR]{near infrared}
\acro{NSBH}[NSBH]{neutron star\nobreakdashes--black hole}
\acro{NSBH}[NSBH]{\acl{NS}\nobreakdashes--\acl{BH}}
\acro{NSF}[NSF]{National Science Foundation}
\acro{NSNS}[NSNS]{\acl{NS}\nobreakdashes--\acl{NS}}
\acro{NS}[NS]{neutron star}
\acro{O1}[O1]{\acl{aLIGO}'s first observing run}
\acro{O2}[O2]{\acl{aLIGO}'s second observing run}
\acro{O3}[O3]{\acl{aLIGO}'s and \acl{AdVirgo}'s third observing run}
\acro{oLIB}[\acrolowercase{o}LIB]{Omicron+\acl{LIB}}
\acro{OT}[OT]{optical transient}
\acro{P48}[P48]{Palomar 48~inch Oschin telescope}
\acro{P60}[P60]{robotic Palomar 60~inch telescope}
\acro{P200}[P200]{Palomar 200~inch Hale telescope}
\acro{PC}[PC]{photon counting}
\acro{PESSTO}[PESSTO]{Public ESO Spectroscopic Survey of Transient Objects}
\acro{PSD}[PSD]{power spectral density}
\acro{PSF}[PSF]{point-spread function}
\acro{PS1}[PS1]{Pan\nobreakdashes-STARRS~1}
\acro{PTF}[PTF]{Palomar Transient Factory}
\acro{QUEST}[QUEST]{Quasar Equatorial Survey Team}
\acro{RAPTOR}[RAPTOR]{Rapid Telescopes for Optical Response}
\acro{REU}[REU]{Research Experiences for Undergraduates}
\acro{RMS}[RMS]{root mean square}
\acro{ROTSE}[ROTSE]{Robotic Optical Transient Search}
\acro{S5}[S5]{\ac{LIGO}'s fifth science run}
\acro{S6}[S6]{\ac{LIGO}'s sixth science run}
\acro{SAA}[SAA]{South Atlantic Anomaly}
\acro{SHB}[SHB]{short, hard burst}
\acro{SHGRB}[SHGRB]{short, hard \acl{GRB}}
\acro{SKA}[SKA]{Square Kilometer Array}
\acro{SMT}[SMT]{Slewing Mirror Telescope\acroextra{ (instrument on \acs{UFFO} Pathfinder)}}
\acro{SNR}[S/N]{signal\nobreakdashes-to\nobreakdashes-noise ratio}
\acro{SSC}[SSC]{synchrotron self\nobreakdashes-Compton}
\acro{SDSS}[SDSS]{Sloan Digital Sky Survey}
\acro{SED}[SED]{spectral energy distribution}
\acro{SFR}[SFR]{star formation rate}
\acro{SGRB}[SGRB]{short \acl{GRB}}
\acro{SN}[SN]{supernova}
\acroplural{SN}[SN\acrolowercase{e}]{supernova}
\acro{SNIa}[\acs{SN}\,I\acrolowercase{a}]{Type~Ia \ac{SN}}
\acroplural{SNIa}[\acsp{SN}\,I\acrolowercase{a}]{Type~Ic \acp{SN}}
\acro{SNIcBL}[\acs{SN}\,I\acrolowercase{c}\nobreakdashes-BL]{broad\nobreakdashes-line Type~Ic \ac{SN}}
\acroplural{SNIcBL}[\acsp{SN}\,I\acrolowercase{c}\nobreakdashes-BL]{broad\nobreakdashes-line Type~Ic \acp{SN}}
\acro{SVD}[SVD]{singular value decomposition}
\acro{TAROT}[TAROT]{T\'{e}lescopes \`{a} Action Rapide pour les Objets Transitoires}
\acro{TDOA}[TDOA]{time delay on arrival}
\acroplural{TDOA}[TDOA\acrolowercase{s}]{time delays on arrival}
\acro{TD}[TD]{time domain}
\acro{TOA}[TOA]{time of arrival}
\acroplural{TOA}[TOA\acrolowercase{s}]{times of arrival}
\acro{TOO}[TOO]{target\nobreakdashes-of\nobreakdashes-opportunity}
\acroplural{TOO}[TOO\acrolowercase{s}]{targets of opportunity}
\acro{UFFO}[UFFO]{Ultra Fast Flash Observatory}
\acro{UHE}[UHE]{ultra high energy}
\acro{UVOT}[UVOT]{UV/Optical Telescope\acroextra{ (instrument on \emph{Swift})}}
\acro{VHE}[VHE]{very high energy}
\acro{VISTA}[VISTA@ESO]{Visible and Infrared Survey Telescope}
\acro{VLA}[VLA]{Karl G. Jansky Very Large Array}
\acro{VLT}[VLT]{Very Large Telescope}
\acro{VST}[VST@ESO]{\acs{VLT} Survey Telescope}
\acro{WAM}[WAM]{Wide\nobreakdashes-band All\nobreakdashes-sky Monitor\acroextra{ (instrument on \emph{Suzaku})}}
\acro{WCS}[WCS]{World Coordinate System}
\acro{WSS}[w.s.s.]{wide\nobreakdashes-sense stationary}
\acro{XRF}[XRF]{X\nobreakdashes-ray flash}
\acroplural{XRF}[XRF\acrolowercase{s}]{X\nobreakdashes-ray flashes}
\acro{XRT}[XRT]{X\nobreakdashes-ray Telescope\acroextra{ (instrument on \emph{Swift})}}
\acro{ZTF}[ZTF]{Zwicky Transient Facility}
\end{acronym}

%% file: table_1_data.tex
MS200615h & 2020-06-15 00:35:40 & 2.02e-06  & -2.9 & -1.9 & 0.1 & 7.1& \url{https://gcn.gsfc.nasa.gov/gcn3/27951.gcn3} \\
MS200618aq & 2020-06-18 05:47:05 & 1.78e-07  & -53.1 & -52.1 & -50.1 & -35.1& \url{https://gcn.gsfc.nasa.gov/gcn3/27990.gcn3} \\
MS200618bq & 2020-06-18 11:00:59 & 3.50e-06  & -16.9 & -21.9 & -11.9 & -2.9& \url{https://gcn.gsfc.nasa.gov/gcn3/27987.gcn3} \\
MS200618bx & 2020-06-18 12:17:08 & 3.76e-06  & -63.3 & -62.3 & -59.3 & -51.3& \url{https://gcn.gsfc.nasa.gov/gcn3/27988.gcn3} \\
MS200619bf & 2020-06-19 10:24:43 & 1.91e-06  & -41.0 & -40.0 & -35.0 & -27.0& \url{https://gcn.gsfc.nasa.gov/gcn3/27989.gcn3} \\